# A deep learning model for segmentation of geographic atrophy to study its long-term natural history

Bart Liefers[1,2*], MSc, Johanna M. Colijn[3,4], MD, MSc, Cristina González-Gonzalo[1,2], MSc, Timo Verzijden[3,4], MSc, Paul Mitchell[5], MD, PhD, Carel B. Hoyng[2,6], MD, PhD, Bram van Ginneken[1], PhD, Caroline C.W. Klaver[2,3,4,6,7], MD, PhD, Clara I. Sánchez, PhD[1,2,6]

[1] Diagnostic Image Analysis Group, Department of Radiology, Radboud University Medical Center, Nijmegen, The Netherlands

[2] Donders Institute for Brain, Cognition and Behaviour, Radboud University Medical Center, Nijmegen, The Netherlands

[3] Department of Ophthalmology, Erasmus University Medical Center, Rotterdam, The Netherlands

[4] Department of Epidemiology, Erasmus University Medical Center, Rotterdam, The Netherlands

[5] Centre for Vision Research, Department of Ophthalmology, The Westmead Institute for Medical Research, The University of Sydney, Sydney, NSW, Australia

[6] Department of Ophthalmology, Radboud University Medical Center, Nijmegen, The Netherlands.

[7] Institute for Molecular and Clinical Ophthalmology, Basel, Switzerland

[*] Corresponding author.
*Address*: Diagnostic Image Analysis Group, Department of Radiology and Nuclear Medicine, Radboud University Medical Center, Geert Grooteplein 10
6525 GA Nijmegen, The Netherlands.
*E-mail address*: Bart.Liefers@radboudumc.nl

**Financial support:** The Rotterdam Study is funded by Erasmus Medical Center and Erasmus University, Rotterdam, Netherlands Organization for the Health Research and Development (ZonMw), the Research Institute for Diseases in the Elderly (RIDE), the Ministry of Education, Culture and Science, the Ministry for Health, Welfare and Sports, the European Commission (DG XII), and the Municipality of Rotterdam. The ophthalmic research within the Rotterdam Study was supported by Oogfonds, Landelijke Stichting voor Blinden en Slechtzienden, Novartis Foundation and MaculaFonds that contributed through UitZicht (grants 2015-36). Other funding was obtained from the AMI project, a collaborative project of the Fraunhofer Gesellschaft and the Radboud University and University Medical Center. The sponsor or funding organization had no role in the design or conduct of this research.

**Conflict of Interest:** No conflicting relationship exists for any author

**Author contributions:**

Research design: Bart Liefers, Caroline C.W. Klaver, Clara I. Sánchez
Data acquisition: Bart Liefers, Johanna M. Colijn, Timo Verzijden, Caroline C.W. Klaver, Clara I. Sánchez
Data analysis: Bart Liefers, Johanna M. Colijn, Cristina González-Gonzalo, Paul Mitchell, Carel B. Hoyng, Bram van Ginneken, Caroline C.W. Klaver, Clara I. Sánchez
Manuscript Preparation: Bart Liefers, Johanna M. Colijn, Timo Verzijden, Caroline C.W. Klaver, Clara I. Sánchez

**Running head:** Deep learning for segmentation of GA

**Online Supplemental Materials:** This article contains additional online-only material. The following should appear online-only: Figures S1, Table S1, Deep learning model details

**Abbreviations and Acronyms: GA** = geographic atrophy; **AMD =** age-related macular degeneration; **CFI** = color fundus image; **OCT** = optical coherence tomography; **FAF** = fundus autofluorescence; **BMES** = Blue Mountains Eye Study; **RS** = Rotterdam Study; **AREDS** = Age-Related Eye Disease Study;

## Abstract

**Purpose:** To develop and validate a deep learning model for the automatic segmentation of geographic atrophy (GA) in color fundus images (CFIs) and its application to study growth rate of GA.

**Design:** Prospective, multicenter, natural history study with up to 15 years of follow up.

**Participants:** 409 CFIs of 238 eyes with GA from the Rotterdam Study (RS) and the Blue Mountain Eye Study (BMES) for model development, and 5,379 CFIs of 625 eyes from the Age-Related Eye Disease Study (AREDS) for analysis of GA growth rate.

**Methods:** A deep learning model based on an ensemble of encoder-decoder architectures was implemented and optimized for the segmentation of GA in CFIs. Four experienced graders delineated, in consensus, GA in CFIs from RS and BMES. These manual delineations were used to evaluate the segmentation model using 5-fold cross-validation. The model was further applied to CFIs from the AREDS to study the growth rate of GA. Linear regression analysis was used to study associations between structural biomarkers at baseline and GA growth rate. A general estimate of the progression of GA area over time was made by combining growth rates of all eyes with GA from the AREDS set.

**Main Outcome Measures:** Automatically segmented GA and GA growth rate.

**Results:** The model obtained an average Dice coefficient of 0.72 ± 0.26 on the BMES and RS set while comparing the automatically segmented GA area to the graders' manual delineations. An intraclass correlation coefficient of 0.83 was reached between the automatically estimated GA area and the graders' consensus measures. Eight automatically calculated structural biomarkers (area, filled area, convex area, convex solidity, eccentricity, roundness, foveal involvement and perimeter) were significantly associated with growth rate. Combining all growth rates indicated that GA area grows quadratically up to an area of around 12 mm², after which growth rate stabilizes or decreases.

**Conclusion**: The presented deep learning model allowed for fully automatic and robust segmentation of GA in CFIs. These segmentations can be used to extract structural characteristics of GA that predict its growth rate.

# Introduction

Geographic atrophy (GA) occurs in the advanced stage of age-related macular degeneration (AMD). It is characterized by progressive atrophy of the retinal pigment epithelium, overlying photoreceptors, and underlying choriocapillaris.[1] Areas of GA often initially appear extrafoveal, where they may cause difficulties in reading or dim-light vision.[2] Over time the atrophic area may grow, and when it reaches the fovea, visual acuity is severely diminished. Prevalence of GA increases exponentially with age,[3] and is highest in people of European ancestry.[4] The number of people affected by GA is expected to increase further in the near future because of the ageing population.[5]

Currently, no approved treatment exists to prevent progression of GA.[6,7] However, several potential therapies are in clinical trial.[8] For evaluation of these trials, reliable anatomic endpoints are required, as visual acuity alone provides insufficient insight in the severity of the disease.[9] Growth rate of the atrophic area has been suggested as an important indicator of disease progression.[9-11] However, the speed at which GA progresses varies greatly between subjects.[12-14] Therefore, understanding the patterns associated with progression and the variability between subjects is important for the design and interpretation of clinical trials.

To assess growth rate, accurate delineation of the GA area is required. However, as manual delineation can be challenging and time-consuming,[15,16] automatic segmentation could provide a scalable and reproducible alternative. Deep learning has emerged as a powerful technique for the automatic analysis of medical images.[17] Deep learning models require labeled examples (training data) to tune their internal parameters. The model then learns to extract features that are important for the segmentation task without further need of explicit domain knowledge from experts. It has been applied successfully to color fundus images (CFIs) for classification of severity stages in AMD[18,19] or diabetic retinopathy,[20] and recently also for the detection of GA.[21] Although manually labeled examples are still required for training and validation, the model can thereafter be applied to large data sets without further intervention of expert ophthalmologists.

These automatic methods also have the potential to efficiently and accurately extract structural characteristics of GA as seen in imaging that have been demonstrated to correlate with growth rate. For example, multifocal lesions grow faster than unifocal lesions[22] and extrafoveal lesions grow faster than foveal lesions.[13] Circular lesions have been demonstrated to grow at a slower rate than more irregularly shaped lesions.[23]. Baseline lesion area has been consistently associated with future growth, with larger lesions growing faster than smaller lesions.[11,13,24,25] However, applying a square root transformation to the lesion size may remove this dependency.[16,26] It is therefore hypothesized that lesions with approximate circular shape grow at a constant radial speed, thus leading to a quadratic growth of the area.[16,27]

Various imaging modalities have been used to assess GA. CFIs are historically most widely used, particularly in large epidemiologic studies.[12] More recently fundus autofluorescence (FAF) and optical coherence tomography (OCT) have also become popular for the study of GA and GA progression.[13,16,25] Several lesion characteristics visible on those modalities can be linked to progression of GA. For example, banded or diffuse perilesional patters on FAF and structural abnormalities at the junctional zone on OCT have been associated with faster GA progression.[13,25,28] Although GA may be detected earlier on FAF than CFI,[29] good agreement on quantification of GA area in CFI between two independent reading centers has been demonstrated,[11] and progression rates assessed from both FAF and CFI are highly correlated.[13,29] CFI has the advantage that it is widely available, often over longer time periods, making it suitable for the study of long term progression of GA. Previous work on automatic methods for segmentation of GA focuses mainly on OCT[30-32] or FAF.[33] Feeny et al.[34] proposed a method based on a random forest classifier in CFI. In contrast, in this study we present a model that is based on deep learning. To our best knowledge, this is the first deep learning model for segmentation of GA in CFI.

The purpose of this study is two-fold: 1) to develop and validate a fully automatic model for segmentation of GA in CFIs. and 2) to demonstrate its utility in a longitudinal setting for the study of GA progression. The performance of the developed model is compared against four graders on a challenging dataset to evaluate its robustness. Next, the automatically segmented GA areas provide measures of structural characteristics related to lesion size, location and morphology. We investigate the associations between those structural characteristics at baseline and subsequent growth rate of GA. Finally, we combine GA growth rates across patients to obtain an estimate of average progression of GA area over time.

## Methods

### Data

Data for development and evaluation of the deep learning model for GA segmentation were collected from the Blue Mountains Eye Study (BMES)[35] and the Rotterdam Study (RS) cohorts I, II and III.[36] The developed model was applied to CFIs from the Age-Related Eye Disease Study (AREDS)[11] for the assessment of GA growth rate.

The BMES is a population study from the Blue Mountains region in Australia that started between 1992 and 1994, and included 3,654 participants aged 49 or older. CFIs were obtained with a Zeiss fundus camera (Carl Zeiss, Oberkochen, Germany) for the first 3 visits and a CanonCF-60 DSi with DS Mark II body (Canon, Tokyo, Japan) for the 4th visit. The BMES was approved by the University of Sydney and the Sydney West Area Health Service Human Research Ethics Committees.

The RS is a population study from a suburb in Rotterdam, the Netherlands. RS cohort I started in 1990 and included 7,983 participants aged 55 years and older. Cohort II started in 2000 and included 3,011 participants aged 55 years and older. Cohort III started in 2006 and included 3,932 participants aged 45 years and older. CFIs for the first examinations were obtained with a Topcon TRV-50VT (Topcon Optical Company, Tokyo, Japan), the last two examinations with a Topcon TRC 50EX and a Sony DXC-950P digital camera. The RS was approved by the Medical Ethics Committee of the Erasmus MC and by the Netherlands Ministry of Health, Welfare and Sport.

The AREDS is a long-term, multicenter, prospective study of the clinical course of AMD and cataract. Starting between 1992 and 1998, 11 clinics in the United States enrolled 4,757 participants aged between 55 and 80 years. Stereoscopic CFIs were acquired with a Zeiss FF-series camera (Carl Zeiss AG, Oberkochen, Germany). The AREDS was approved by an independent institutional review board at each clinical center.

The follow up interval for RS and BMES was five years. The AREDS had follow up at six months intervals, although the typical interval between available CFIs was one year. The BMES, RS and AREDS all adhere to the tenets of the Declaration of Helsinki.

A total of 504 CFIs of patients diagnosed with AMD and signs of GA were included from the BMES and RS sets. 26 images with mixed signs of AMD (neovascularization, bleedings, scars) were excluded in order to disambiguate overlapping areas. Furthermore, no GA was delineated in 43 images because it was either not present or ungradable, and 26 images were excluded due to poor image quality. The remaining 409 images were included for development of the model and evaluation of its performance.

This set contains 87 images from BMES (26 participants, 43 eyes) and 322 images from RS (149 participants, 195 eyes). The 409 images represent 315 unique visits (some visits had two CFIs available).

We identified 5,379 images (459 participants, 625 eyes) from the AREDS set with GA and at least two years of follow up, following the grading available from the database of genotype and phenotype (dbGaP) 2014 table. Most of these images were stereoscopic, so this accounted for 2,750 unique acquisitions (eye-visit). Pixel to millimeter conversion was fixed for all images, based on the average distance between fovea and center of the optic disc measured in a subset of the images. This distance was assumed to be 4.5mm.[37]

Delineations of GA area were made by four graders (3 of them with over 20 years of experience), using an in-house created software platform for manual annotations (https://www.a-eyeresearch.nl/software/ophthalmology_workstation/).[38] For RS, additional multimodal imaging (infrared, FAF and/or OCT) was available for some of the visits, and the platform allowed images of the same eye (both multimodal and longitudinal) to be aligned manually by identifying corresponding landmarks. The graders could simultaneously view images of the same eye using a synchronized cursor on multiple screens. GA was identified as absence of the retinal pigment epithelium and increased visibility of the choriocapillaris on CFI. Additional evidence from other modalities was used whenever available. Areas of macular and peripapillary atrophy were delineated as separate classes, but for this study only macular GA was used.

Each grader annotated the entire BMES set, while the RS set was divided in such a way that each grader annotated approximately half of the entire set and every image was graded by at least two graders. Finally, a consensus grading was made for all images in both sets. During the consensus grading all graders decided together which of the individual gradings was most accurate, and updated this grading if necessary, until consensus was reached. If two CFIs of the same visit were present, both were included for model development and the delineated GA area was propagated from one image to the other by using the affine transformation calculated from the manual landmarks. For evaluation, only the CFI that was used to make the consensus grading was used.

## Model

The proposed deep learning model for GA segmentation consisted of an ensemble of several models, each trained with partly overlapping training sets. The network architecture (the topology of connections between internal parameters of the deep-learning model) for each model consisted of a deep encoder-decoder structure with residual blocks and shortcut connections, similar to De Fauw et al,[39] but adapted to work with CFIs. This architecture, and its variations, can be characterized by a contracting path, in which the high-resolution input image is converted to a low-resolution abstract representation, followed by an expanding path in which the original resolution is reconstructed. The contracting and expanding path are connected by shortcut connections. This approach has been shown to be very effective for semantic segmentation in medical imaging for which large contextual information is required.

Input to each model was both the original color image and a contrast-enhanced version of the same image, both resampled to 512x512 pixels. The contrast-enhanced image was obtained by subtracting a blurred image from the original image.[40] The input was transformed through the many layers of artificial neurons in the contracting and expanding path, and ultimately yielded a new image in which the value of every pixel represented a likelihood of being part of an area of GA. A threshold was applied to this likelihood image to obtain the final GA area. More details about the model and the training procedure used for this study can be found in the supplementary material.

## GA segmentation

For the development and the validation of the model, we applied a five-fold cross-validation scheme. Data from BMES and RS were merged into one dataset and randomly split at patient level into five approximately equal folds. In a rotating scheme, four folds were used for model training and validation (development set), while the remaining fold was used for performance evaluation (test set). Furthermore, four separate models were created within each development set. Each model used three folds for tuning of the internal parameters (training) and one for validation. An ensemble of these four models was then evaluated on the respective test set. Ultimately, an ensemble of the 20 obtained models (four models developed for each of the five rounds) constituted the final model.

The performance of the model and the agreement between graders were assessed using the Dice coefficient, which is defined as two times the intersection of two areas divided by the sum of the individual areas. Hence, a value of zero represents disjoint areas (no overlap), while a value of one represents perfect agreement. Dice coefficients were calculated between graders to assess the inter-observer agreement, whereas the areas delineated in the consensus grading were used as reference for the model. Note that the consensus grading was not independent of the individual gradings, and therefore could not be used as a reference to estimate graders' performance. Furthermore, intra-class correlation coefficient (ICC) of the GA area and of the square root of the GA area was used to measure agreement between graders and the model.

## GA growth rate

The final deep learning model (the ensemble of 20 models) was applied to CFIs from AREDS for the analysis of GA progression. It is well-documented that GA area increases faster for larger lesions. To remove the dependency of baseline lesion size on growth rate, many researchers apply a square root transformation to the GA area.[26] Similarly, we calculated the square root annual growth in millimeter per year for each eye to assess progression in the AREDS set.[37] This value was obtained from the slope of a linear regression through the square root of the GA area for a selected set of timepoints. The selected set consisted of all available CFIs within a window of 2 years, for which the number of available CFIs was highest for the respective eye. The window was limited at 2 years because growth rate and lesion characteristics may change over time.[23] We calculated the correlation of square root annual growth rate between fellow eyes, and compared growth rate between groups using an unpaired t-test for unilateral versus bilateral, unifocal versus multifocal and foveal versus extrafoveal cases.

In order to identify structural characteristics or features that may be predictive for growth rate, we built a linear model based on features that were extracted from the segmented GA area at baseline (the first image within the selected window). Candidate features were area, perimeter, convex area, filled area, solidity (area / convex area ratio and area / filled area ratio), number of lesions, eccentricity, circularity, roundness and foveal involvement. Details on how these features were calculated can be found in the supplementary material. Associations between individual features and square root annual growth rate were calculated using univariate linear regression. Because the features were not independent, a multivariate linear model was created to further investigate which features best explain variation in square root annual growth rate. The multivariate model was built using forward selection, by iteratively adding the feature that yielded the highest increase in adjusted $R^2$ value, until it no further increased. When stereoscopic images were available, lesion characteristics were represented by the mean of the two calculated values. In order to obtain a more homogeneous set for the prediction model, we discarded images where the relative difference in GA area between the left and right stereoscopic image was more than 50%, and only included eyes with at least 2 years of follow up images.

Finally, we combined all estimates of GA growth in a single figure. GA growth in mm² per year (not square root transformed) was estimated as a function of GA area, again using a linear regression for each eye through the GA area in a window of 2 years. This resulted in an estimate of GA growth (the slope of the regression), bounded by a minimum and maximum GA area. The estimated general GA growth for a given GA area was then represented by the mean of all growth estimates for which this GA area fell within the respective area bounds. Confidence intervals were estimated using bootstrapping.

## Results

### GA segmentation

The deep learning model reached a Dice coefficient of 0.72 ± 0.26 (N=315), measured in cross-validation in the BMES and RS data sets. Dice coefficients between two independent graders ranged from 0.72 ± 0.26 to 0.82 ± 0.21 (0.78 ± 0.24 on average). See Table 1 for more details. The intraclass correlation coefficient between the model and the consensus was 0.83 for GA area, and 0.84 for the square root of the GA area. Consistency in those values is further visualized in Figure 1 using Bland-Altman plots. The mean value of the differences between consensus and model did not differ significantly from 0 on the basis of a 1-sample t-test for neither GA area (p=0.82) nor square root GA area (p=0.22). Examples of manually and automatically segmented GA areas can be found in Figure 2. More examples of automatic segmentation results on the AREDS set can be found in Supplementary Figure 1.

### GA growth rate

After excluding visits where the difference between left and right stereoscopic images in automatically segmented area was more than 50%, 584 of the 625 eyes in AREDS with at least 2 years of follow up remained. Square root annual growth of GA for those eyes was 0.21 ± 0.46 mm/year. This value was significantly higher for eyes with small (<5 mm²) baseline GA area (0.31 ± 0.39, N=308), compared to eyes with large (≥ 5mm²) baseline GA area (0.10 ± 0.50, N=276), p<0.001. Table 2 shows differences in growth rate between groups. We observed that multifocal and extrafoveal lesions grow faster than unifocal or foveal lesions. Subjects with bilateral GA showed faster progression than unilateral cases, although not significant in our analysis (p=0.12). Growth rates between fellow eyes were correlated (r=0.45, p<0.001). Figure 3 highlights progression of GA for selected individual eyes.

Correlations between baseline lesion characteristic and square root annual growth are summarized in Table 3. Eight out of eleven features were significantly correlated with GA growth rate (after Bonferroni correction). Features included in the multivariate model were area, circularity, filled solidity, convex area, number of lesions, eccentricity and roundness. The coefficient of determination of this model was 0.18.

A visualization that summarizes growth over time for all eyes with GA in the AREDS set can be found in Figure 4. The red dashed line in these graphs represent a quadratic model that best fitted the data for GA area < 12 mm².

## Discussion

A deep-learning model for segmentation of GA in CFIs was developed and evaluated. We demonstrated how the automatically obtained segmentations of the model can be used to study growth rate of GA on an independent set. The performance of the deep learning model in terms of Dice coefficient on the BMES and RS set approached that of human experts. The model was able to identify GA even when image quality or contrast were relatively poor, as demonstrated in Figure 2. Nevertheless, some failure cases were still present, which was the main reason for the lower average Dice coefficient. We suspect that more training data may solve this issue, since each of the models only used 60% of the data (~245 images) for training, which may not be enough given the inherent difficulty of the problem and the variability in the data. For application to the AREDS set this problem was partly circumvented by using an ensemble model, which indirectly made use of all training data.

The obtained mean square root annual growth rate on the AREDS set (0.21 ± 0.46 mm/year) was slightly lower than previously reported values. For example, Domalpally et al. observed 0.30 mm/year,[29] and Keenan et al. observed 0.28 mm/year.[41] A reason for this may be the dependence of growth rate on baseline area. When we split the dataset on baseline lesion size, we observed that small lesions have larger square root growth rates (see Table 2). This phenomenon was analyzed in more detail in Figure 4. A quadratic curve seemed to fit the observed GA progression very well up to an area of around 12 mm². For larger areas, the growth rate seemed to stabilize or even decrease. Similar observations were made by Keenan et al.[41], whose reported values are included in Figure 4 for comparison.

The importance of baseline area for assessing growth rate also became apparent in the regression analysis, where area, filled area and convex area were most strongly correlated with square root annual growth rate. However, when we included only lesions with baseline area < 12mm² in the regression analysis, no features related to lesion size were significantly associated with square root annual growth rate. On an individual level, we also observed a quadratic growth of the area of GA in many cases in the AREDS set, some of them highlighted in Figure 3, where we fitted a quadratic curve through the GA area over time. Again, the decrease in growth rate for larger lesions was visible (bottom two cases in Figure 3).

Of the features that are invariant to lesion size, convex solidity was most significantly associated with square root annual growth rate. Convex solidity is low for irregular shaped lesions, but also for multifocal lesions. This feature hence captures multiple previously reported associations. Circularity was previously associated with GA growth rate,[23] but compared to other features, the association was not very strong in our analysis. An explanation is that the model may have produced a segmentation with a very jagged border for some lesions with indistinct borders of the atrophic area. This could have led to a relatively large perimeter, and hence a lower value for circularity. Roundness will be a better representation of how well the lesion approaches a circular shape in those cases, as it represents the ratio of the area of an enclosing circle and the area of the lesion, and is hence less sensitive to irregular borders.[42]

A limitation of our study was that the conversion from pixels to millimeters may have been inaccurate. This conversion was based on the average distance between fovea and optic disk in a subset of images. Although it is unlikely that this inaccuracy was a source for bias in reported associations with growth rate, reported values for area and growth rate may be slightly larger or smaller in reality.

In the future, we will extend the model to other modalities, specifically FAF and OCT. This may give more accurate measurements of the atrophic area, and hence more reliable assessment of growth rate. In this study, only morphological features of the atrophic area were considered. A next step would be to include associations between growth rate and other lesions patterns, especially those visible on FAF or OCT. Finally, we are investigating the capabilities of deep learning models to directly predict areas where GA may develop. This will provide predictions of both the extent and the location of future GA area.

In conclusion, we have presented and validated a robust segmentation model based on deep learning for GA in CFIs. The model was capable of reproducing known associations between current GA status and future growth. Moreover, we indicated novel structural biomarkers that are predictive for future growth rate, such as solidity, eccentricity or roundness of the lesion. We demonstrated how deep learning can help in the automation of grading, allowing for analysis of larger datasets and helping to understand progression of GA.

## Acknowledgements

Manual delineations of GA in the BMES and RS set were performed by the EyeNED Reading Center, specifically by Johanna Colijn, Caroline Klaver, Corina Brussee and Ada Hooghart.

**Table 1:** Dice coefficients between model and consensus grading, and between individual graders.

|                     | N   | Dice coefficient |
|---------------------|-----|------------------|
| Model – Consensus   | 315 | 0.72 ± 0.26      |
| Grader 1 – Grader 2 | 146 | 0.80 ± 0.27      |
| Grader 1 – Grader 3 | 138 | 0.78 ± 0.27      |
| Grader 1 – Grader 4 | 90  | 0.72 ± 0.26      |
| Grader 2 – Grader 3 | 91  | 0.82 ± 0.21      |
| Grader 2 – Grader 4 | 134 | 0.78 ± 0.22      |
| Grader 3 – Grader 4 | 130 | 0.78 ± 0.19      |

**Table 2:** Square root annual growth of the GA area. Values represent mean ± standard deviation. P-values are calculated using an unpaired t-test.

|  | Square root annual growth (mm/year) | | |
| --- | --- | --- | --- |
|  | All | Small (<5mm$^2$) | Large (≥5mm$^2$) |
| overall | 0.21 ± 0.46 (N:584) | 0.31 ± 0.39 (N:308) | 0.10 ± 0.50 (N:276) |
| unifocal | 0.18 ± 0.41 (N:401) | 0.26 ± 0.33 (N:201) | 0.10 ± 0.46 (N:200) |
| multifocal | 0.28 ± 0.54 (N:183) | 0.39 ± 0.47 (N:107) | 0.12 ± 0.59 (N:76) |
| P-value | 0.017 | 0.005 | 0.784 |
| foveal | 0.18 ± 0.45 (N:459) | 0.28 ± 0.35 (N:192) | 0.10 ± 0.50 (N:267) |
| extrafoveal | 0.33 ± 0.44 (N:125) | 0.35 ± 0.44 (N:116) | 0.12 ± 0.28 (N:9) |
| P-value | 0.001 | 0.131 | 0.900 |
| unilateral | 0.18 ± 0.47 (N:217) | 0.26 ± 0.36 (N:115) | 0.10 ± 0.54 (N:102) |
| bilateral | 0.23 ± 0.45 (N:367) | 0.32 ± 0.41 (N:193) | 0.14 ± 0.47 (N:174) |
| P-value | 0.123 | 0.386 | 0.174 |

**Table 3:** Correlations between baseline lesion characteristics (features) and square root annual growth rate (in mm/year). Features are sorted in decreasing order of strength of association. A P-value smaller than 0.0045 (0.05, Bonferroni corrected) is considered significant

| Feature | $R^2$ | Slope | Intercept | R | P | Std error |
|---|---|---|---|---|---|---|
| Area | 0.115 | -0.016 | 0.365 | -0.339 | <0.001 | 0.002 |
| Filled area | 0.114 | -0.015 | 0.365 | -0.338 | <0.001 | 0.002 |
| Convex area | 0.097 | -0.013 | 0.363 | -0.312 | <0.001 | 0.002 |
| Convex solidity | 0.086 | -0.814 | 0.891 | -0.294 | <0.001 | 0.132 |
| Eccentricity | 0.069 | 0.677 | -0.194 | 0.263 | <0.001 | 0.124 |
| Roundness | 0.065 | -0.706 | 0.72 | -0.255 | <0.001 | 0.133 |
| Fovea region | 0.045 | -3.045 | 0.373 | -0.213 | <0.001 | 0.695 |
| Perimeter | 0.045 | -0.008 | 0.363 | -0.213 | <0.001 | 0.002 |
| Number of lesions | 0.016 | 0.057 | 0.144 | 0.127 | 0.011* | 0.022 |
| Circularity | 0.014 | -0.227 | 0.338 | -0.119 | 0.017* | 0.095 |
| Filled solidity | 0.004 | 1.445 | -1.198 | 0.062 | 0.212* | 1.154 |

* Not significant

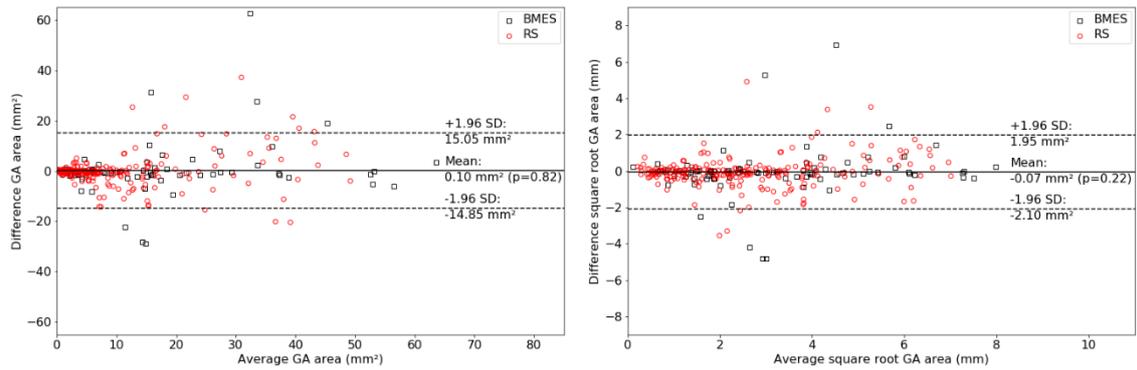

**Figure 1**: Bland-Altman plot of GA area (left) and square root GA area (right). Differences are calculated as the area/square root area of the consensus grading minus the automatic segmentation.

**Figure 2:** Examples of automatic GA segmentation. The green area corresponds to either the consensus (left) or the model output (right). The top three rows show accurate segmentation results, for various configurations of GA differing in area, shape and number of lesions, and variable image quality and contrast. The bottom row shows examples of inaccurate model output.

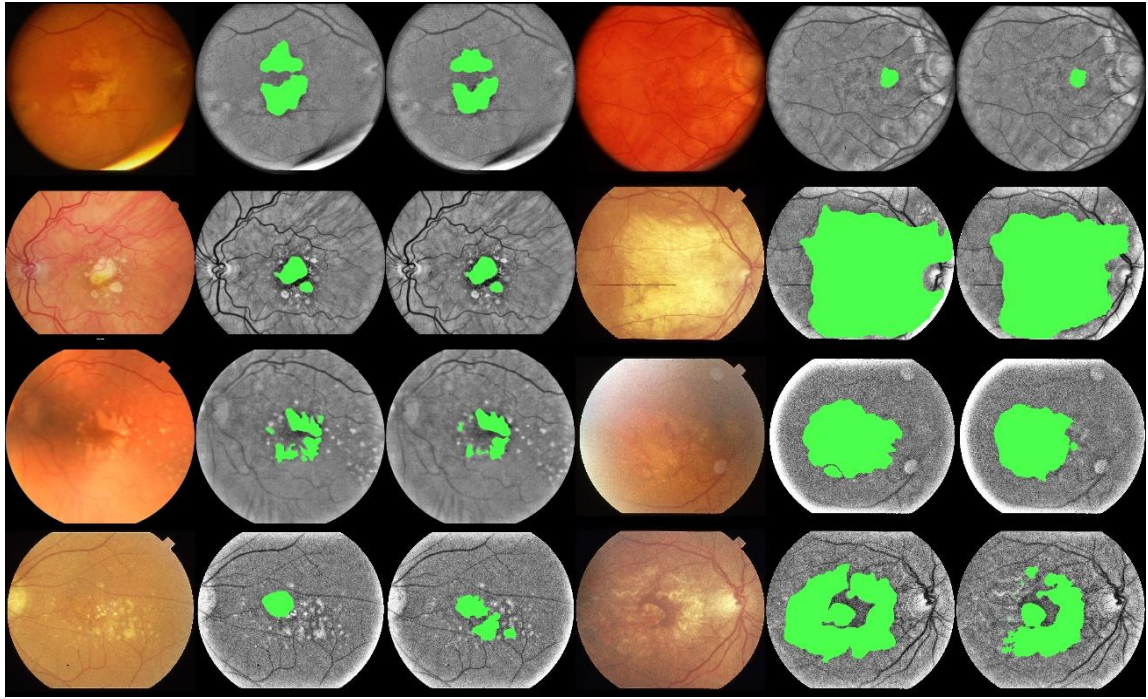

**Figure 3:** Progression of GA over time for 4 selected eyes. The graphs represent area measurements over time (two points per timepoint for the LS and RS stereoscopic images). The blue line is a quadratic fit through the points. For the top 2 cases, an increment in growth rate can be observed. 53834 LE has a more irregular shape than 51551 RE and progresses faster. In the bottom two cases we observe that the growth decreases as the GA area gets larger.

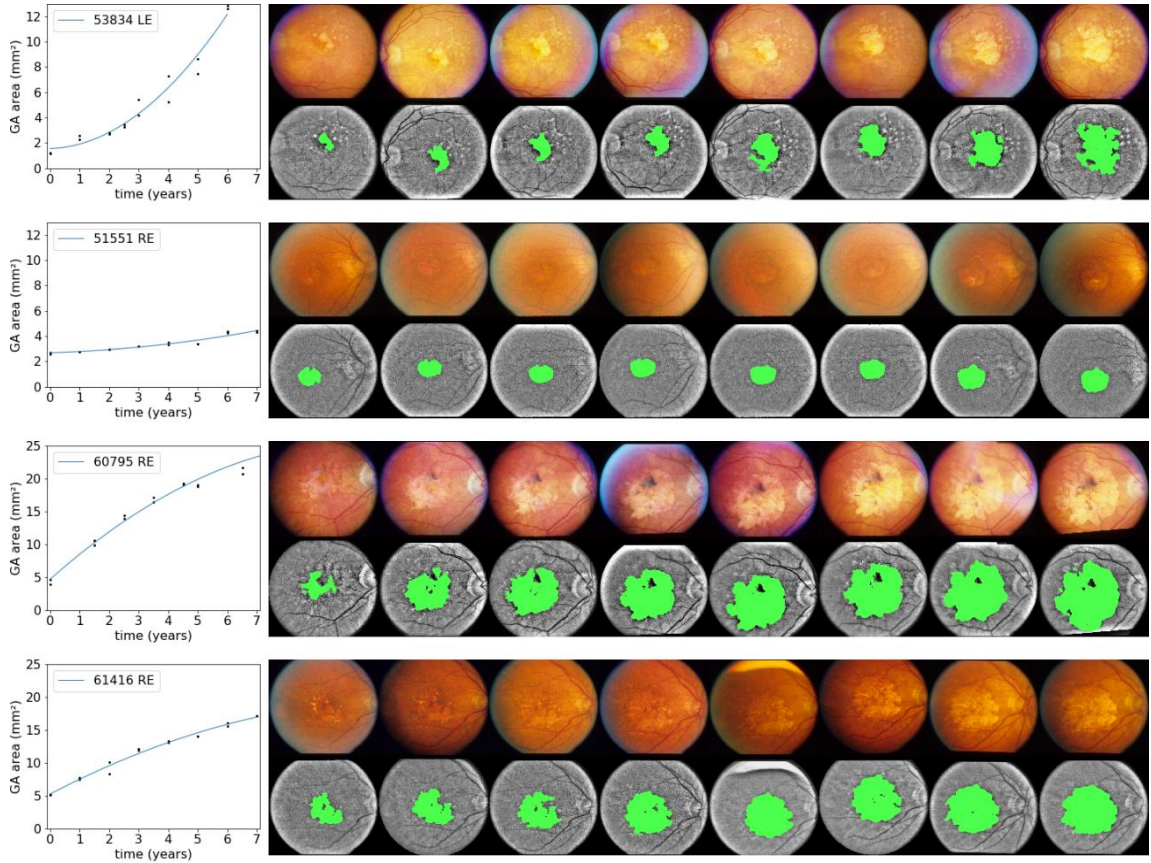

**Figure 4**: GA growth over time. Left: GA growth rate (in mm²/year) as a function of GA area. The blue line represents growth rates estimated from the segmentations of the deep learning model. The shaded area represents the 95% confidence interval (estimated using bootstrapping). The dashed red line represents the growth rate of a quadratic model, as visualized in the right graph. Right: the blue line represents the evolution of GA area over time, obtained by numerically integrating the estimated growth rates from the left graph using a GA area of 0.5 mm² at t=0. The red dashed line represents the best quadratic fit to the plot for GA area < 12 mm². Above this area the observed GA area diverges from the quadratic fit.

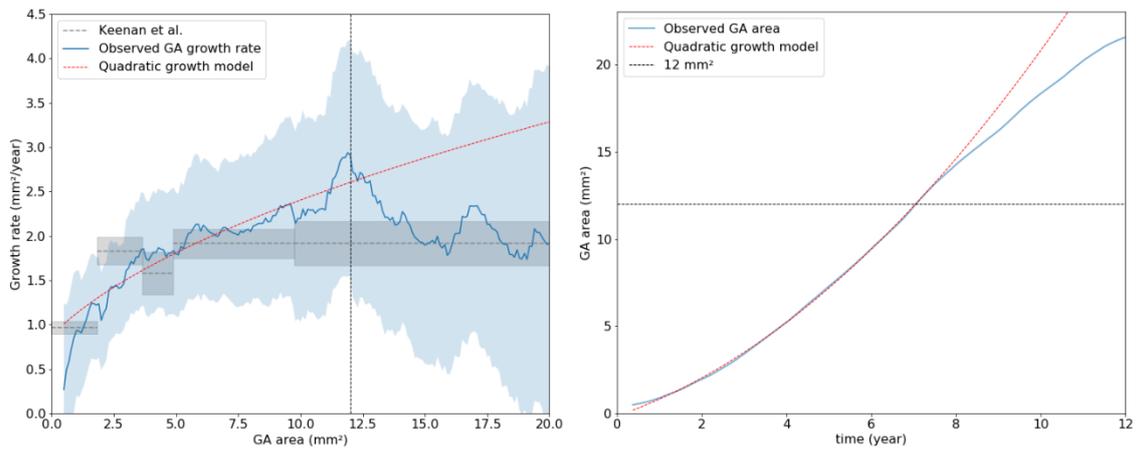

# Appendix

## Deep learning model details

The deep learning architecture consisted of an encoder-decoder structure with eight levels of resolution, connected by shortcut connections at every level. At every level of resolution, a residual block with two 3x3 convolutions was used. The number of filters per convolution was, for each of the respective levels 32, 32, 64, 64, 128, 128, 256, 256. The eight levels of down-sampling reduce the input from 512x512 to a feature map of 2x2 pixels. At the lowest level, two residual blocks with 1x1 convolutions and 2048 filters each were applied. The down-sampling operations were performed by strided convolutions.

Binary cross-entropy was used as loss-function. During the first 10 epochs, the loss was weighted to balance the classes. The model was trained on batches of 2 images, using the adam-optimizer with a learning rate of $32 * 10^{-5}$. The learning rate was divided by two every 50 epochs. Input images were augmented by horizontal and vertical flipping, scaling of up to 1.3, rotations of maximum 40 degrees, and translations of up to 150 pixels.

The best model for each of the subsets in the cross-validation scheme was selected based on best performance on the respective validation set. Performance was assessed as best average Dice-coefficient after selecting the optimal threshold. The ensemble model was constructed by combining the output of the models after correcting for differences in optimal threshold between models:

$$y = \sum_k y_k^{\log(2)/\log(th_k)}$$

Where $y$ represents the final prediction for a specific pixel, $y_k$ represents the prediction for that pixel for model $k$ and $th_k$ represents the optimal threshold for model $k$.

Implementation and training of the model was done using the keras library[1] with tensorflow backend.[2]

1. Chollet, F. et al., Keras, https://keras.io, 2015;
2. Abadi, M. et al. TensorFlow: Large-scale machine learning on heterogeneous systems, https://www.tensorflow.org, 2015

## Description of lesion characteristics

**Area**: The total area of all segmented GA.

**Perimeter**: The total perimeter along the border of the segmented area. If lesions have holes, the perimeter along inner borders are also included. Four-connectivity for border pixel determination is used.

**Convex area**: The area of the convex hull of all lesions: multifocal lesions are joined rather than calculating the convex hull for each focus separately.

**Filled area**: The area of the segmented GA with holes filled.

**Convex solidit**y: The ratio of area and convex area.

**Filled solidity**: The ratio of area and filled area.

**Number of lesions**: The number of separate lesions with a diameter of at least 0.175 mm. Lesions are separated if pixels do not touch neither horizontally, vertically or diagonally.

**Eccentricity**: Eccentricity of the ellipse that has the same second-moments as the GA region. The eccentricity is the ratio of the focal distance (distance between focal points) over the major axis length.

**Circularity**: Calculated as $4\pi$ area / perimeter².

**Roundness**: The ratio of the actual GA area and the area of an enclosing circle, calculated as 4 area / $\pi$ d², where d is the length of the major axis of the ellipse that has the same second central moments as the region.

**Foveal involvement**: The intersection area of the segmented GA area and a circular area with diameter of 0.3 mm in the center of the image.

All lesion characteristics were calculated using Python 3.7, with the numpy library, version 1.15.3 (https://numpy.org/) and the scikit-image library, version 0.14.1 (https://scikit-image.org/)

**Figure S1:** Sample showing the output of the model for the first 20 patients in the AREDS set (showing the first image with label GA for every patient).

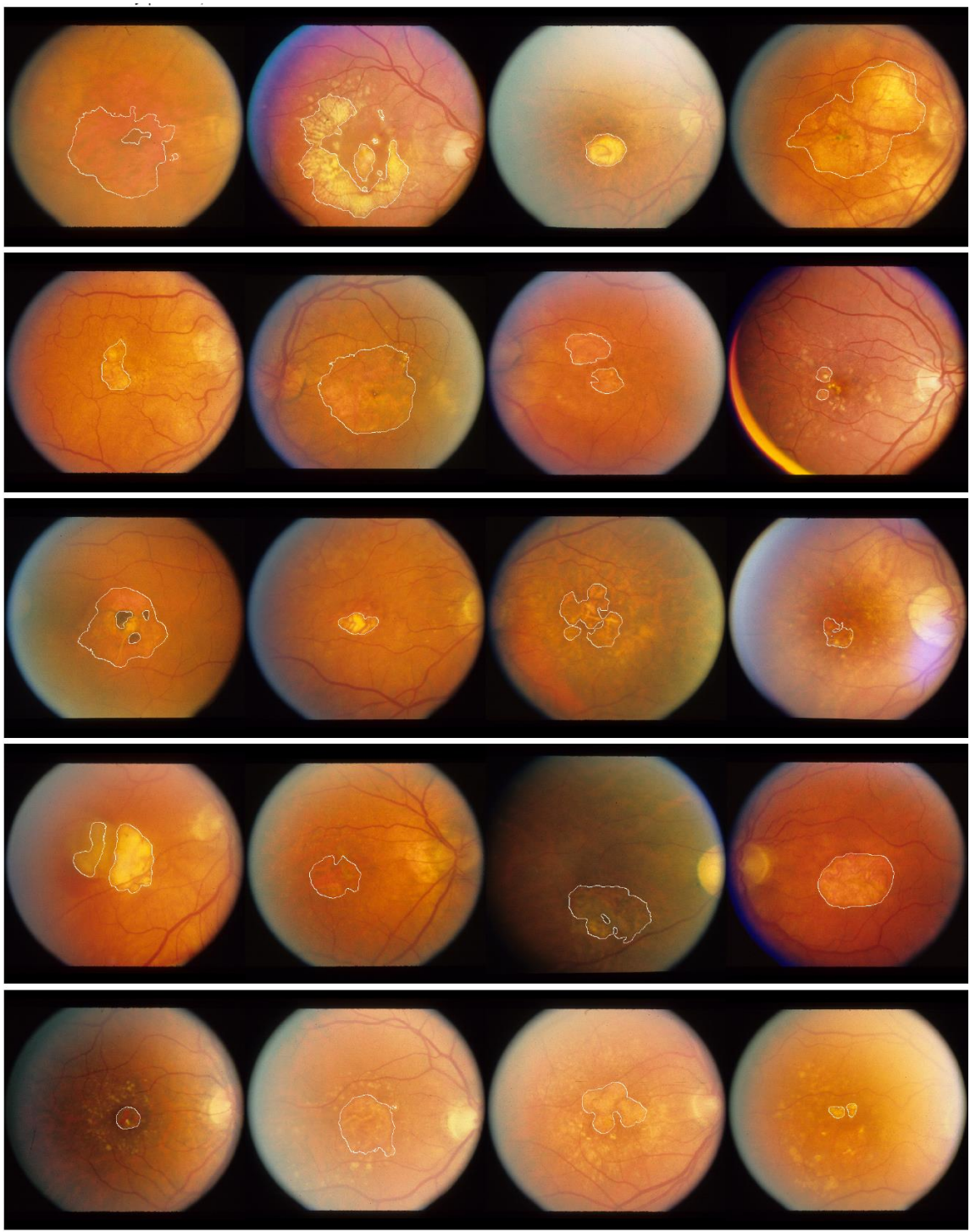